\documentclass[pra,superscriptaddress,reprint,amsmath,amssymb,aps]{revtex4-2}

\usepackage{graphicx}
\usepackage{float}
\usepackage{diagbox}
\usepackage[
pdfstartview=FitH,
bookmarks=true,
breaklinks=true,
colorlinks=true,
linkcolor=blue,anchorcolor=blue,
citecolor=blue,filecolor=blue,
menucolor=blue,urlcolor=blue]{hyperref}

\begin{document}

\title{Calibrating the absorption imaging of cold atoms under high magnetic fields}

\author{Yuqi Liu}
\thanks{These authors contribute equally to this work.}
\author{Zhongchi Zhang}
\thanks{These authors contribute equally to this work.}
\author{Shiwan Miao}
\author{Zihan Zhao}
\author{Huaichuan Wang}
\affiliation{Department of Physics and State Key Laboratory of Low Dimensional Quantum Physics, Tsinghua University, Beijing, 100084, China}
\author{Wenlan Chen}
\email{cwlaser@ultracold.cn}
\author{Jiazhong Hu}
\email{hujiazhong01@ultracold.cn}
\affiliation{Department of Physics and State Key Laboratory of Low Dimensional Quantum Physics, Tsinghua University, Beijing, 100084, China}
\affiliation{Frontier Science Center for Quantum Information, Beijing, 100084, China}
\affiliation{Collaborative Innovation Center of Quantum Matter, Beijing, 100084, China}

\date{\today}

\begin{abstract}
We develop a theoretical model for calibrating the absorption imaging of cold atoms under high magnetic fields. Comparing to zero or low magnetic fields, the efficiency of the absorption imaging becomes lower while it requires an additional correction factor to obtain the absolute atom number under the Beer-Lambert law. 
Our model is based on the rate equations and can account many experimental imperfections such as Zeeman level crossing, failures of hyperfine structures, off-resonant couplings, and low repumping efficiency, etc.
Based on this method, we can precisely calculate the correction factor for atom number measurement without any empirical or fitting parameters. Meanwhile, we use a cold-atom apparatus of rubidium-85 to experimentally verify our model. Besides these, we find our work can also serve as a benchmark to measure the polarization impurity of a circular-polarized laser beam with high sensitivities.
We believe this work will bring convenience for most of cold-atom experiments using absorption imaging.
\end{abstract}

\maketitle
\section{Introduction}

The cold-atom platform has become a thriving area of research in recent decades, offering a powerful new tool for studying many-body physics \cite{MorschRMP2006,BlochRMP2008,Bloch2012,Gross2017} thanks to its unique ability to directly measure the density of atoms in both real and momentum space \cite{Schafer2020,Gross2021}. A plethora of novel phenomena have been observed, including the momentum bimodal distribution in Bose-Einstein condensates \cite{BEC1,BEC2,Stellmer22013}, phonon dispersion \cite{Steinhauer2002,Ernst2010,Guo2021}, and the superfluid-to-Mott transition in optical lattice systems \cite{SFtoMOTT_2002,Gemelke2009,Bakr2010,LIANG20222550}. Absorption imaging is the most commonly used technique to observe these phenomena, where the density distribution of atoms is obtained by detecting the absorption of near-resonant light by the atoms. When atomic transitions are not saturated, the Beer-Lambert law predicts that the transmission of light decreases exponentially with increasing atomic density. By comparing the transmission intensity with and without atoms, we can obtain the density distribution of atoms.

Meanwhile, magnetic fields also play a crucial role in cold atom experiments. By adjusting the magnetic field, we can tune the scattering length via Feshbach resonance~\cite{Feshbach}, where the scattering state energy coincides with a bound molecular state in the close channel. This ability has generated many novel phases and phenomena and has paved the way for new fields
including BEC-BCS crossover \cite{PhysRevLett.92.150402,Gap2004,PhysRevLett.93.050401,Zwierlein2005}, Efimov physics \cite{Kraemer2006,Lompe2010,Pollack2009,Zaccanti2009}, Feshbach dimers \cite{PhysRevLett.92.120403,PhysRevLett.96.030401,Kiefer2023}, degenerate molecules \cite{Zwierlein2003,PhysRevLett.92.120401,Schindewolf2022,Marco2019}, and quench dynamics \cite{hu_quantum_2019,Feng2019,Eigen2018,Zhang2020,PhysRevLett.121.243001,PhysRevLett.127.200601}.
In addition, magnetic fields offer powerful tools beyond tuning the interaction strength, including the use of spin-orbit coupling \cite{Lin2011,PhysRevLett.109.095302,Huang2016,Wu2016,Li2017}, Raman sideband cooling \cite{PhysRevLett.84.439,Hu2017,PhysRevLett.81.5768}, and synthetic gauge fields \cite{Lin2009,PhysRevLett.111.185302,PhysRevLett.111.185301}.

However, absorption imaging is highly sensitive to the energy structure of the atoms \cite{atomicbook}, which can be strongly affected by magnetic fields. As a result of level crossing and lifting of hyperfine level degeneracy, the result of absorption imaging with non-zero magnetic field can be significantly different from that with zero magnetic field due to the low pumping efficiency and changed dipole moments. Nevertheless, with the advancement of quantum simulation, a more-precise measurement of the atom number is required in a growing number of regimes, such as the measurement of equation of state \cite{Yang2017,Yefsah2011,christodoulou2021,Ha2013,hung2011observation} and quantum gas microscope \cite{Gemelke2009,Bakr2010}. Currently, an empirical formula is commonly used to calibrate the absorption imaging by fitting it with experimental data \cite{Yang2017,Yefsah2011}, which may be complicated and limited by the accuracy of the experiments.

In this manuscript, we develop a numerical model to calculate the impact of non-zero magnetic fields on absorption imaging, and we experimentally verify this model by a cold-atom apparatus based on rubidium-85 atoms. 
By taking into account the interplay among varieties of facts including Zeeman level crossing and shifting, the change of dipole moments, light polarization, and repumping efficiency, we can accurately determine the correction factor for the atom number measurement and identify the optimal parameters for imaging purposes. Moreover, this model can also serve as a benchmark for assessing the polarization purity of light since the absorption imaging with some specific apparatus is very sensitive to the polarization.

The manuscript is organized as follows: Section II provides a description of our theoretical model and includes a simplified illustration of how non-zero magnetic fields impact the imaging procedure. Section III presents our experimental demonstration which provides consistent results with our numerical calculations. 
Meanwhile, we give the explanations for each peak and valley on the spectrum.
In Section IV, we show that absorption imaging with non-zero magnetic fields can be utilized as a benchmark for evaluating the quality of circular-polarized light where the sensitivity of the polarization impurity can reach a 0.02\% level. It provides a practical application for inspecting the laser beam in cold-atom apparatus without additional optics or instruments with a high precision. 

\begin{figure}[tb]
\centering
\includegraphics[width=0.46\textwidth]{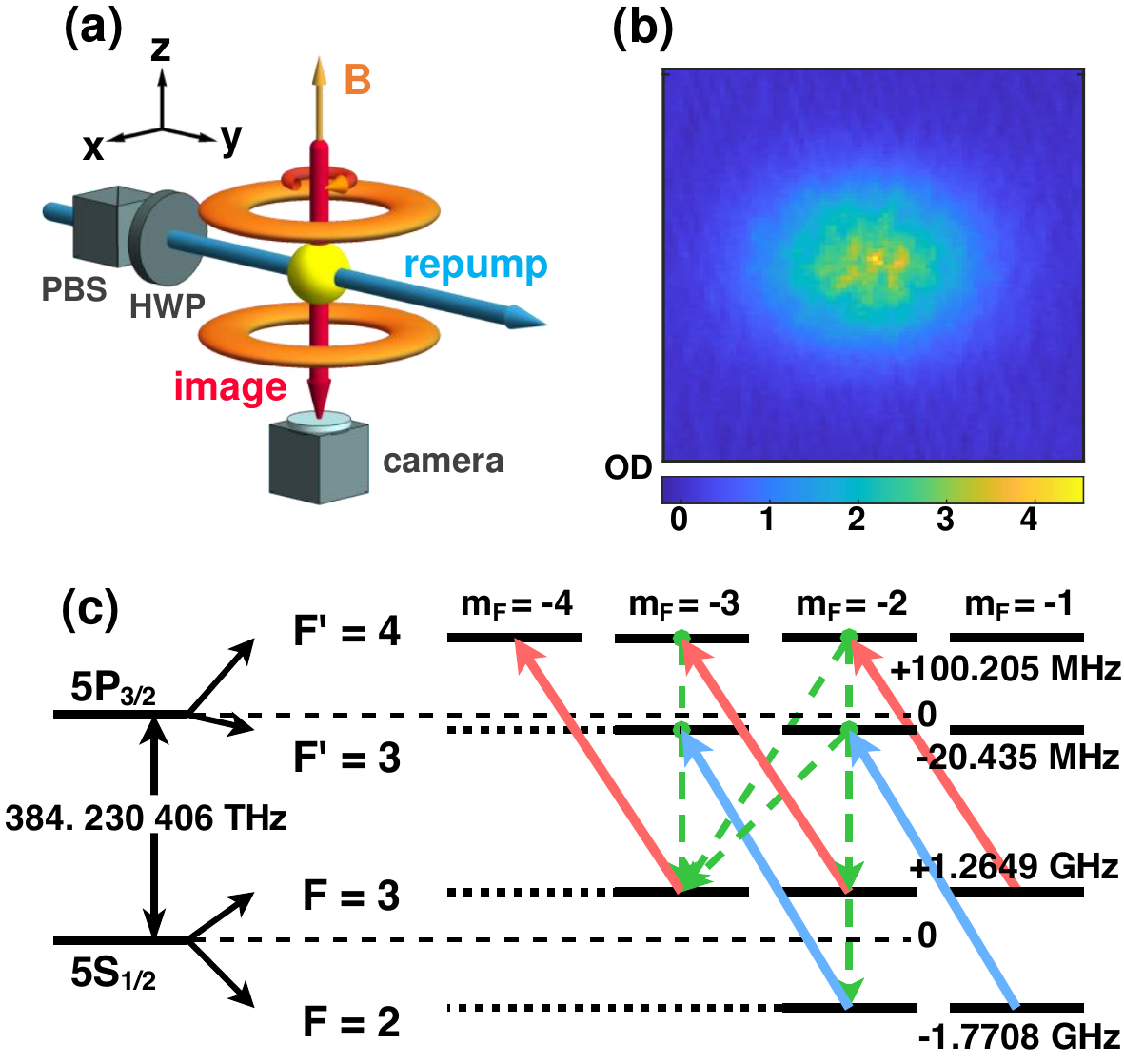}
\caption{\label{Fig1}(a) 
A conventional experimental setup for absorption imaging. 
A pair of Helmholtz coils generate magnetic field along the $z$ axis. The imaging beam is sent into a camera along the $z$ axis with a circular polarization $\sigma^-$.
Besides the imaging beam, a repumping beam is required to clean the other hyperfine levels in ground state manifolds. Here we send the repumping beam along the $y$ axis to avoid it entering the camera. 
(b) An absorption image obtained in our experiment and the optical depth calculated by the Beer-Lambert law. 
(c) Part of energy levels of $^{85}$Rb under zero magnetic field. Red arrows correspond to the imaging light, blue arrows correspond to the repumping light, and green dashed arrows correspond to the spontaneous emission. After a few spontaneous decays, the atomic population accumulates in the state of $5S_{1/2}|3, -3\rangle$ with a closed transition to $5P_{3/2}|4',-4'\rangle$ for the imaging light. }
\end{figure}

\section{Theoretical model for high-field imaging}
In this section, we focus on the theoretical part of high-field imaging and divide the contents into three parts. The first part is a brief summary of the absorption imaging which is the foundation of this work. The second part is a discussion about the challenges of high-field imaging. It explains why people obtain different optical depth or atom number under high magnetic fields. The third part is how we use rate equations to describe the imaging process. Based on these, we calculate the imaging efficiency $\beta$.  

\subsection{A brief summary of absorption imaging}
 First, we will provide a brief summary of absorption imaging. When a beam of light, $I_{in}(x,y)$, propagates parallel to the z-axis and enters a sample of atoms, themodified Beer-Lambert law states that \cite{Hueck:17,Reinaudi:07} 
\begin{equation}
\frac{dI(x,y,z)}{dz}=-n(x,y,z) \sigma \frac{I(x,y,z)}{1+I(x,y,z)/I_{sat}}.
\label{Eq:OD}
\end{equation}
By measuring the laser intensity before and after the sample, the optical density ($OD$), which is the product of the absorption cross section $\sigma$ and the area density $n_{2D}(x,y)=\int_{-\infty}^{+\infty} n(x,y,z)dz$, can be calculated as 
\begin{equation}
OD(x,y)=n_{2D}(x,y) \sigma = -\ln\left[{I_{f}(x,y) \over I_{i}(x,y)}\right]+\frac{I_{i}-I_{f}}{I_{sat}}.
\end{equation}
The absorption cross section, $\sigma$, is defined as
\begin{equation}
\sigma =
\frac{\hbar \omega \Gamma}{2I_{sat}}\frac{1}{1+4(\Delta/\Gamma)^{2}}.
\end{equation}
Here $\Delta=\omega-\omega_0$ is the detuning between the imaging light frequency $\omega$ and the atomic resonance $\omega_0$, $\Gamma$ is the spontaneous emission rate.
The saturation intensity, $I_{sat}$, can be calculated by
\begin{equation}
\label{EqSat}
    I_{sat}
    = {c\epsilon_{0} \Gamma^{2} \hbar^{2} \over 4 |\hat{p} \cdot \mathbf{d}|^{2}},
\end{equation}
where $\epsilon_{0}$ is the vacuum dielectric constant, $c$ is the speed of light, $\hat{p}$ is the unit polarization vector of the light field, and $\mathbf{d}$ is the dipole moment.
In the following paragraphs, we use the atomic total angular momentum, $F$, and its projection on the quantization axis ($z$-axis), $m_{F}$, to label a ground state $|s\rangle = |F,m_{F}\rangle$ or an excited state $|s'\rangle = |F',m'_{F}\rangle$, where the prime indicates an excited state.
The dipole moment of alkali atoms between the state $|s\rangle$ and $|s'\rangle$ can be calculated by the formula:
\begin{equation}
    d_{|s\rangle,|s'\rangle,p} = \alpha_{|s\rangle,|s'\rangle,p}\sqrt{2J+1 \over 2J'+1} \langle J||e\mathbf{r}||J' \rangle,
\end{equation}
where $J$ and $J'$ label the total electronic spin, $\langle J||e\mathbf{r}||J' \rangle$ is the transition dipole matrix element that can be found in data sheets \cite{steck2rubidium}, and $\alpha_{|s\rangle,|s'\rangle,p}$ is a coefficient. The subscript $p$ here denotes the projection into the unit polarization vector $\hat p$ of the light field.

\begin{table}[tb]
\caption{\label{Table1}
The square of the dipole moment coefficients $|\bm{\alpha}|^{2}$ between the manifolds of $5^{2}P_{3/2}$ and $5^{2}S_{1/2}$ in $^{85}$Rb under zero magnetic field.}
\renewcommand\arraystretch{1.2}
\resizebox{\linewidth}{!}{
\begin{tabular}{|c|c|c|c|c|c|}
\hline
\diagbox{$5P_{3/2}$}{$5S_{1/2}$}&$|3,-3\rangle$&$|3,-2\rangle$&$|2,-2\rangle$&$|3,-1\rangle$&$|2,-1\rangle$\\
\colrule
$|4',-4'\rangle$&$1   $&$0    $&$0   $&$0   $&$ 0$\\
\hline
$|4',-3'\rangle$&$1/4 $&$3/4  $&$0   $&$0   $&$ 0$\\
\hline
$|3',-3'\rangle$&$5/12$&$5/36 $&$4/9 $&$0   $&$ 0$\\
\hline
$|4',-2'\rangle$&$1/28$&$3/7 $&$0   $&$15/28$&$0$\\
\hline
$|3',-2'\rangle$&$5/36$&$5/27 $&$4/27$&$25/108$&$7/18$\\
\hline
$|2',-2'\rangle$&$10/63$&$10/189$&$14/27$&$2/189$&$7/27$\\
\hline
$|4',-1'\rangle$&$0   $&$3/28 $&$0   $&$15/28$&$ 0$\\
\hline
$|3',-1'\rangle$&$0   $&$25/108$&$4/135$&$5/108$&$32/135$\\
\hline
$|2',-1'\rangle$&$0   $&$20/189$&$7/27$&$16/189$&$7/54$\\
\hline
$|1',-1'\rangle$&$0   $&$0$&$3/5$&$0$&$3/10$\\
\hline
\end{tabular}
}
\end{table}

For better illustration, we have chosen rubidium-85 atoms to demonstrate the energy levels in detail. This choice does not affect the generality of our results.
In Fig.~1, we present a conventional experimental setup with the energy level structure.
We also list the square of the dipole moment coefficient, $|\alpha_{|s\rangle,|s'\rangle}|^{2}=|\sum_{p}\alpha_{|s\rangle,|s'\rangle,p}|^{2}$, for some transitions of $^{85}$Rb under zero magnetic field. 
Typically, a closed transition is employed to maximize the photon scattering of the imaging beam. In this transition, the excited state can only decay into one ground state. 
For $^{85}$Rb, the closed transition is the transition between $5S_{1/2} |3,-3\rangle$ and $5P_{3/2} |4',-4'\rangle$ for an imaging beam with a $\sigma^-$ polarization.

When the magnetic field is small such the Zeeman splitting is less than the spontaneous emission rate, we can ignore the change of light detunings between the ground and excited states due to Zeeman shift.
All atoms in the $F=3$ manifold can be excited by the imaging light with a $\sigma^-$ polarization (red arrows in Fig.~1(c)), which also connects the closed transition $5S_{1/2}|3,-3\rangle\rightarrow 5P_{3/2}|4',-4'\rangle$. 
However, atoms may also initially occupy the ground states with $F=2$, so a repumping beam (blue arrows) is required to resonantly pump atoms from the $F=2$ manifold to the $F=3$ manifold via the excited states with $F'=3$.
After several spontaneous emissions, all atoms will be initialized into the closed transition with $5S_{1/2}|3,-3\rangle$, regardless of any initial populations in ground states.

\subsection{Difficulties of imaging under high magnetic field}

Under high magnetic fields, the Zeeman energy shift becomes larger than the spontaneous emission rate and the total angular momentum $F$ describing the hyperfine structure is no longer a good quantum number. In Particular, the $5P_{3/2}$ manifold of excited states are strongly mixed together by magnetic fields since the hyperfine splitting for excited states is much smaller than that of ground states. Therefore, we need to re-diagonalize the eigenstates under both magnetic fields and hyperfine interactions.

\begin{figure}[tb]
\centering
\includegraphics[width=0.46\textwidth]{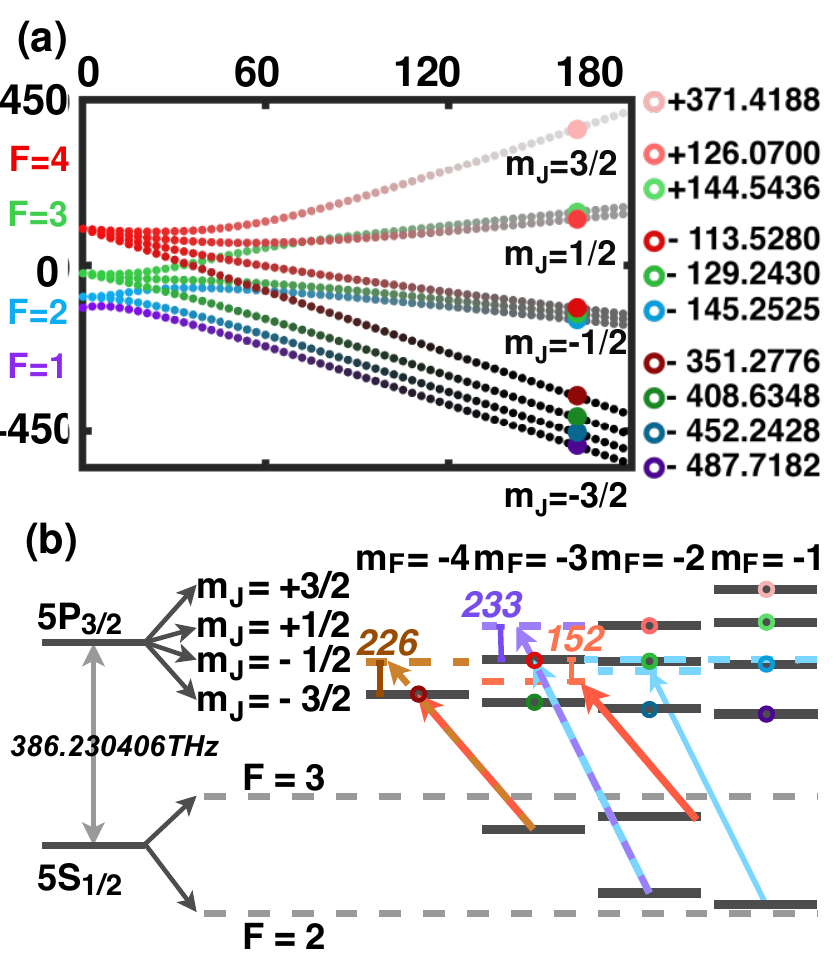}
\caption{\label{Fig2}(a) 
The energy levels of $^{85}$Rb $5P_{3/2}$ state were plotted for different magnetic fields. The angular momentum along the $z$-axis of the displayed levels varies from -4 to -1. 
The filled circles mark the levels at 161~G magnetic field with the energy values listed on the right in the units of MHz.  
(b) Atomic structure at 161~G. Red solid arrows and blue solid arrows respectively represent the imaging light and repumping light. The brown dashed arrows and purple dashed arrows correspond to the imaging light and repumping light, respectively, in the absence of magnetic field. It should be noted that the imaging light and repumping light require red-detuning by 226~MHz and 233~MHz to adapt to the magnetic field. Specifically, the imaging light has a red detuning of 152~MHz for the transition between $5^{2}S_{1/2}|3,-2\rangle$ and $5^{2}P_{3/2}|4',-3'\rangle$.}
\end{figure}

In the absence of a magnetic field, we use the hyperfine structure $|F,m_F\rangle$ to identify the eigenstates. However, at non-zero magnetic fields, neither $F$ nor $|F,m_F\rangle$ are good quantum numbers. To label the states that are adiabatically connected to the zero-field state $|F,m_F\rangle$, we use the notation $|\Tilde{F},\Tilde{m_F}\rangle$. At very high magnetic fields, where the Zeeman splitting greatly exceeds the hyperfine splitting, the states $|\Tilde{F},\Tilde{m_F}\rangle$ are actually labeled by another set of good quantum numbers $|m_J,m_I\rangle$ where $m_J$ and $m_I$ are the angular momenta of electrons and nuclei along the magnetic field. 
We have plotted the energy levels versus the magnetic field for the $5P_{3/2}$ manifold in Fig.~\ref{Fig2}(a). The dipole moments betweeen different states are also affected by the magnetic field, and we have provided the square of dipole moment coefficients ($|\alpha|^2$) of $^{85}$Rb under 161~G magnetic field in Table~\ref{Table2} for comparison.

As the magentic field increases, 
the dipole moments between different states are strongly distorted and this has a significant impact on the frequencies of repumping lasers that we choose.
For examples, the transition $5P_{3/2} |\Tilde{3'},\Tilde{-3'}\rangle\rightarrow 5S_{1/2} |\Tilde{3},\Tilde{-3}\rangle$ is strongly suppressed due to different nuclear spins at high field.
However, some dipole transitions that are forbidden under zero magnetic field become no longer forbidden, such as the transition $5S_{1/2} |\Tilde{2},\Tilde{-2}\rangle\rightarrow 5P_{3/2} |\Tilde{4'},\Tilde{-3'}\rangle$.
Meanwhile, the closed transition $5P_{3/2} |\Tilde{4}',\Tilde{-3}'\rangle\rightarrow 5S_{1/2} |\Tilde{3},\Tilde{-3}\rangle$ is not affected by the magnetic field since the hyperfine states involved in this transition do not change with respect to magnetic field.
Therefore, under high magnetic field, we should set the repumping light resonant with the transition $5S_{1/2} |\Tilde{2},\Tilde{-2}\rangle\rightarrow 5P_{3/2} |\Tilde{4'},\Tilde{-3'}\rangle$ instead of $5P_{3/2}|\Tilde{3'},\Tilde{-3'}\rangle$ to initialize most of the atoms into the state of $5S_{1/2}|\Tilde{3},\Tilde{-3}\rangle$ (Fig.~\ref{Fig2}).
This is a significant difference in repumping between zero or high magnetic fields.

\begin{table}[tb]
\caption{\label{Table2}
The square of the dipole moment coefficients $|\bm{\alpha}|^{2}$ between manifolds $5P_{3/2}$ and $5S_{1/2}$ of $^{85}$Rb under 161~G magnetic field. We label $5P_{3/2}$ states by both $|m_{J}',m_{I}'\rangle$ and $|\Tilde{F'},\Tilde{m_{F}'}\rangle$.}
\renewcommand\arraystretch{1.5}
\resizebox{\linewidth}{!}{
\begin{tabular}{|c|c|c|c|c|c|}
\hline
\footnotesize\diagbox{$5P_{3/2}$\\ $|\Tilde{F'},\Tilde{m_{F}}'\rangle$\\($|m_{J}',m_{I}'\rangle$)}{$5S_{1/2}$\\$|\Tilde{F},\Tilde{m_{F}}\rangle$}&$|\Tilde{3},-\Tilde{3}\rangle$&$|\Tilde{3},-\Tilde{2}\rangle$&$|\Tilde{2},-\Tilde{2}\rangle$&$|\Tilde{3},-\Tilde{1}\rangle$&$|\Tilde{2},-\Tilde{1}\rangle$\\
\hline
$|\Tilde{4'},-\Tilde{4'}\rangle$ \normalsize$\left(\left|-{3\over 2}',-{5\over 2}'\right\rangle\right)$
&$1 $&$0   $&$0  $&$0  $&$ 0$\\
\hline
$|\Tilde{4'},-\Tilde{3'}\rangle$ \normalsize$\left(\left|-{1\over 2}',-{5\over 2}'\right\rangle\right)$ &$0.6394$&$0.1945$&$0.1660$&$0$&$ 0$\\
\hline
$|\Tilde{3'},-\Tilde{3'}\rangle$ \normalsize$\left(\left|-{3\over 2}',-{3\over 2}'\right\rangle\right)$ &$0.0272$&$0.6624$&$0.3104$&$0$&$0$\\
\hline
$|\Tilde{4'},-\Tilde{2'}\rangle$ \normalsize$\left(\left|+{1\over 2}',-{5\over 2}'\right\rangle\right)$ &$0.3163$&$0.2804$&$0.3852$&$0.0125$&$0.0056$\\
\hline
$|\Tilde{3'},-\Tilde{2'}\rangle$ \normalsize$\left(\left|-{1\over 2}',-{3\over 2}'\right\rangle\right)$ &$0.0170$&$0.3648$&$0.2743$&$0.2566$&$0.0873$\\
\hline
$|\Tilde{2'},-\Tilde{2'}\rangle$ \normalsize$\left(\left|-{3\over 2}',-{1\over 2}'\right\rangle\right)$ &2.3E-05&0.0215&$0.0071$&$0.4617$&$0.5097$\\
\hline
$|\Tilde{4'},-\Tilde{1'}\rangle$ \normalsize$\left(\left|+{3\over 2}',-{5\over 2}'\right\rangle\right)$&$0 $&$0.2952$&$0.6840$&$0.0126$&$0.0077$\\
\hline
$|\Tilde{3'},-\Tilde{1'}\rangle$ \normalsize$\left(\left|+{1\over 2}',-{3\over 2}'\right\rangle\right)$&$0 $&$0.1657$&$0.1678$&$0.4195$&$0.2259$\\
\hline
$|\Tilde{2'},-\Tilde{1'}\rangle$ \normalsize$\left(\left|-{1\over 2}',-{1\over 2}'\right\rangle\right)$ &$0$&$0.0155$&$0.0051$&$0.2218$&$0.4232$\\
\hline
$|\Tilde{1'},-\Tilde{1'}\rangle$ \normalsize$\left(\left|-{3\over 2}',+{1\over 2}'\right\rangle\right)$ &$0$&1.3E-05&2.2E-07&$0.0128$&$0.0099$\\
\hline
\end{tabular}}
\end{table}

However, the absorption imaging efficiency is still lower than the condition under zero magnetic field. 
It is because some of the atoms pumped from $5S_{1/2} |\Tilde{2},-\Tilde{2}\rangle$ to $5P_{3/2} |\Tilde{4'},-\Tilde{3'}\rangle$ will decay to $5S_{1/2} |\Tilde{3},-\Tilde{2}\rangle$ instead of $ 5S_{1/2}|\Tilde{3},-\Tilde{3}\rangle$. 
In the absence of magnetic field, the degeneracy of the $F=3$ manifold ensures that atoms in $5S_{1/2}|\Tilde{3},\Tilde{-2}\rangle$ will be pumped by the imaging beam and ultimately decay to the state $ 5S_{1/2}|\Tilde{3},\Tilde{-3}\rangle$. 
However, under high magnetic field, atoms scattering to $5S_{1/2}|\Tilde{3},\Tilde{-2}\rangle$ can hardly be pumped to the $5P_{3/2} |\Tilde{4'},-\Tilde{3'}\rangle$ state due to the large detuning (152~MHz at 161~G in Fig.~\ref{Fig2}(b)). As a result, only atoms at $5S_{1/2} |\Tilde{3},-\Tilde{3}\rangle$ can be imaged by the imaging light, leading to a reduction in the absorption imaging efficiency. 
To account for this, an additional coefficient (or efficiency) $\beta$ ($\beta<1$) is required to correct the optical depth and accurately reflect the real atom number in the Beer-Lambert law, which can be expressed as
\begin{equation}
    n(x,y) \sigma \beta= OD(x,y)
\end{equation}
Therefore, accurately predicting and calculating how the efficiency $\beta$ changes with respect to the mangetic field strength is crucial for high-field imaging.

\begin{figure*}[tb]
\centering
\includegraphics[width=0.8\textwidth]{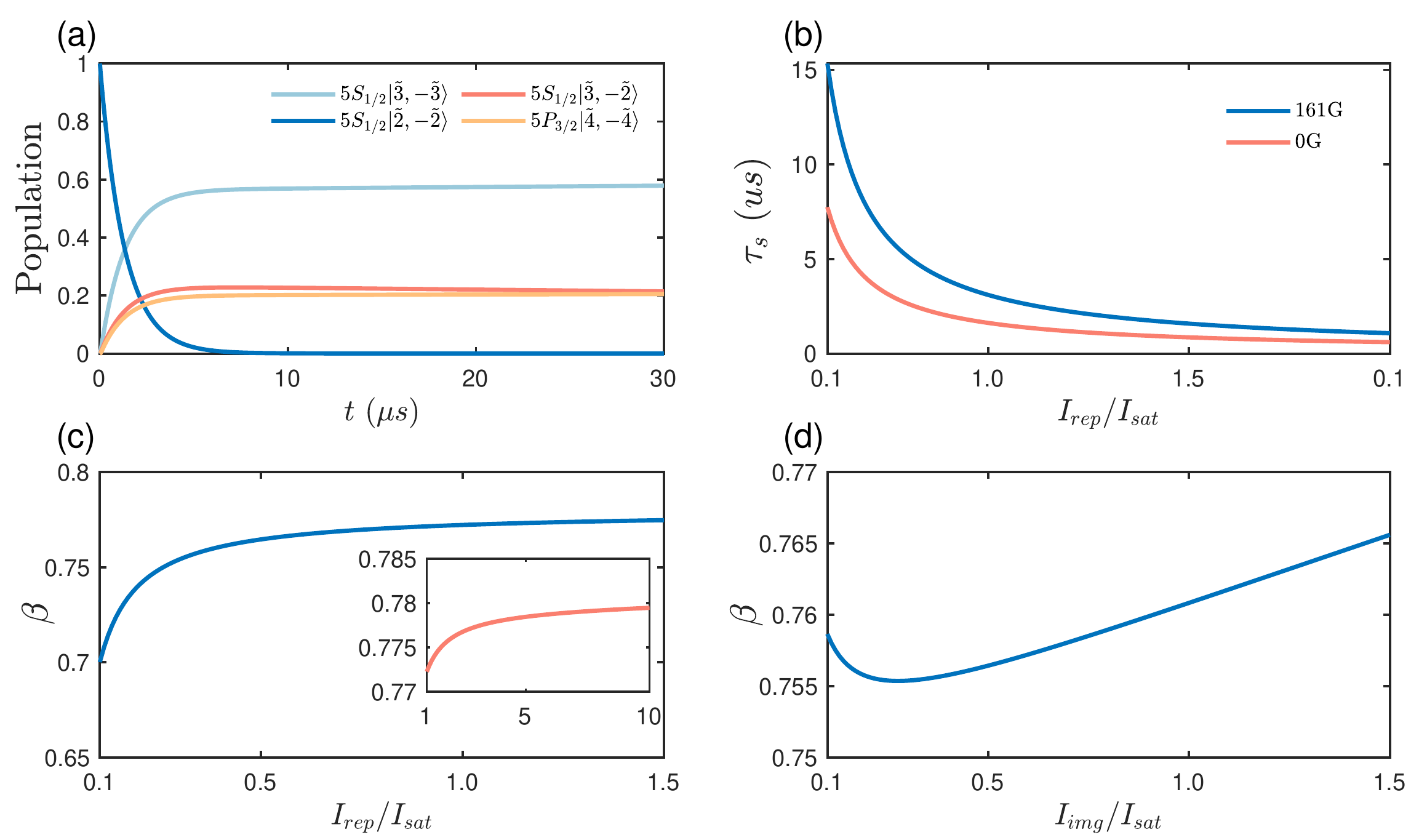}
\caption{\label{Fig3}(a) The temporal evolution of the populations of four main states, namely $5S_{1/2} |\Tilde{3},\Tilde{-3}\rangle$, $5S_{1/2} |\Tilde{3},-\Tilde{2}\rangle$, $5S_{1/2} |\Tilde{2},-\Tilde{2}\rangle$, and $5P_{3/2} |\Tilde{4'},-\Tilde{4'}\rangle$, under the conditions of a magnetic field of 161~G, $I_{img}=I_{sat}$ and $I_{rep}=0.4I_{sat}$.  
(b) The settling time $\tau_s$ versus the repumping light intensity $I_{rep}$ for both 0~G and 161~G magnetic fields. 
(c) The imaging efficiency $\beta$ versus the repumping light intensity $I_{rep}$ under magnetic field of 161~G and an imaging duration of 40~$\mathrm{\mu}$s . The main plot shows the range from 0.1 to 1.5$I_{sat}$, and the inset shows the range from 1 to 10$I_{sat}$. When the repumping light intensity reaches the saturation intensity, the efficency $\beta$ experiences a negligible change of less than 1\%.
(d) The efficiency $\beta$ versus the imaging light intensity $I_{img}$ under magnetic field of 161~G and an imaging duration of 40~$\mathrm{\mu}$s. The change of $\beta$ is below 1.4\%.
}
\end{figure*}

\subsection{Rate equations describing high-field imaging}

In this section, we aim to develop a model for calculating the efficiency $\beta$ of absorption imaging.
According to the definition of $\beta$, we can calculate it by the following approach:
\begin{equation}
\beta =d N_{l}/dN_{l,0G}.  \label{beta} 
\end{equation}
Here $dN_{l}$ is the mean photon number scattered by one atom under a particular magnetic field in a time duration $d\tau$, and $dN_{l,0G}$ is the mean photon number scattered by the same atom under the same conditions, except that there is no magnetic field. 
Our model simulates the process in which an ensemble of atoms is illuminated by a repumping light with intensity $I_{rep}$ and an imaging light with intensity $I_{img}$. We use a vector
 \begin{equation}
 \mathbf{v}_{oc}=\{f_{|F',m'_{F}\rangle},...,f_{|F,m_{F}\rangle},...\}^{T}      
 \end{equation}
to list the population fraction of all energy levels related to the imaging process. Here, $f_{|F',m'_{F}\rangle}$ is the fraction of atoms occupying the state $|F',m'_{F}\rangle$ in the ensemble, and the superscript $T$ means transposed matrix. The sum of all occupation fractions $f_{|s\rangle}$ and $f_{|s'\rangle}$ is 1. The variation of $\mathbf{v}_{oc}$ obeys a rate equation
\begin{equation}
\frac{d\mathbf{v}_{oc}}{dt} = \mathbf{M}_{T} \mathbf{v}_{oc}.\label{Eq3}
\end{equation}
To obtain the exact form of this rate equation, we write down each equation for both excited and ground states with forms of
\begin{eqnarray}
\label{EqRateground}\frac{d f_{|s\rangle}}{dt} = &-&\sum_{l,j,p}\!\left( {\Gamma_{|s'_{j}\rangle}\over 2 }\right)\!{|\alpha_{p,|s\rangle,|s'_{j}\rangle}|^2 I_{l,p}/I_{sat}\over 1+4(\Delta_{l,|s\rangle,|s'_{j}\rangle}/\Gamma_{|s'_{j}\rangle})^{2}}\!(f_{|s\rangle}-f_{|s'_{j}\rangle})\nonumber \\
&+&\  \sum_{j} \Gamma_{|s'_{j}\rangle} |\boldsymbol{\alpha}_{|s\rangle,|s'_{j}\rangle}|^2 f_{|s'_{j}\rangle},\ \mathrm{and}\\
\label{EqRateexcite}\frac{d f_{|s'\rangle}}{dt} = &+&\sum_{l,i,p}\!\left( {\Gamma_{|s'\rangle}\over 2 }\right)\!{|\alpha_{p,|s\rangle,|s'_{j}\rangle}|^2I_{l,p}/I_{sat}\over 1+4(\Delta_{l,|s_{i}\rangle,|s'\rangle}/\Gamma_{|s'\rangle})^{2}}\!(f_{|s_{i}\rangle}-f_{|s'\rangle})\nonumber \\
&-&\  \Gamma_{|s'\rangle} f_{|s'\rangle}. 
\end{eqnarray}
Here $l$ is used to label different beams in the system. $I_{sat}$ represents the saturation intensity calculated using Eq.~\ref{EqSat} with a specific substitution $|\hat{p} \cdot \mathbf{d}_{|s\rangle,|s'\rangle}| = \sqrt{2J+1 \over 2J'+1} \langle J||e\mathbf{r}||J' \rangle$.
$I_{l,p}$ is the  $p$-polarized intensity of the $l$-th beam, and $\Delta_{l,|s\rangle,|s'\rangle}$ is the detuning between the $l$-th light and the transition $|s\rangle\rightarrow |s'\rangle$.
The average photon number $d N_{l}$ scattered by one atom at time $t$ in a time duration $d \tau$ is
\begin{eqnarray}
\frac{d N_{l}(t)}{d\tau} &=& \sum_{|s\rangle,|s'\rangle,p} \left( {\Gamma_{|s'\rangle}\over 2 }\right)\!{\alpha_{|s\rangle,|s'\rangle,p}^{2} I_{l,p}/I_{sat,ref}\over 1+4(\Delta_{l,|s\rangle,|s'\rangle}/\Gamma_{|s'\rangle})^{2}} \nonumber \\
  &\times&  \left(f_{|s\rangle}(t)-f_{|s'\rangle}(t)\right).
\end{eqnarray}
The vector $\mathbf{v}_{oc}$ and transition matrix $\mathbf{M}_{T}$ are written in the partitioned form,  
\begin{eqnarray}in
\mathbf{v}_{oc}&=&\{\mathbf{v}_{|s'\rangle},\mathbf{v}_{|s\rangle}\}^{T},\\
\mathbf{M}_{t}&=&\left[\begin{array}{cc}
\mathbf{M}_{s's'}&\mathbf{M}_{sti}\\
\mathbf{M}_{sti}^{T}+\mathbf{M}_{spon}&\mathbf{M}_{ss}
\end{array}
\right],
\end{eqnarray}
where $\mathbf{v}_{|s'\rangle}$ and $\mathbf{v}_{|s\rangle}$ describe the population fraction of the excited and ground manifolds respectively. 
According to Eq.~\ref{EqRateground} and Eq.~\ref{EqRateexcite}, the sub-matrices of $\mathbf{M}_{T}$ can be expressed as follows:   
\begin{eqnarray}
\left[\mathbf{M}_{sti}\right]_{|s'_{j}\rangle,|s_{i}\rangle} &=& \sum_{l,p}\left({\Gamma_{|s'_{j}\rangle}\over 2 }\right)\!{|\alpha_{p,|s\rangle,|s'_{j}\rangle}|^2 I_{l,p}/I_{sat}\over 1+4(\Delta_{l,|s_{j}\rangle,|s'_{i}\rangle}/\Gamma_{|s'_{i}\rangle})^{2}},\nonumber\\
\left[\mathbf{M}_{spon}\right]_{|s_{i}\rangle,|s'_{j}\rangle} &=& \Gamma_{|s'_{j}\rangle}|\boldsymbol{\alpha}_{|s\rangle,|s'_{j}\rangle}|^{2},\nonumber\\
\left[\mathbf{M}_{ss}\right]_{|s_{i}\rangle,|s_{k}\rangle} &=& -\delta_{ik} \sum_{j} \left[\mathbf{M}_{sti}\right]_{|s'_{j}\rangle,|s_{i}\rangle},\ and\nonumber\\
\left[\mathbf{M}_{s's'}\right]_{|s'_{j}\rangle,|s'_{k}\rangle} &=& -\delta_{jk} \sum_{k} \left[\mathbf{M}_{sti}\right]_{|s'_{j}\rangle,|s_{k}\rangle} - \delta_{jk}\Gamma_{|s'_{j}\rangle},
\end{eqnarray}
where $\delta_{ik}$ is a Kronecker delta symbol. 
Based on rate equations, we calculate $dN_l/d\tau$ for atoms with and without magnetic field, which leads to the coefficient $\beta$ in Eq.~(\ref{beta}). 
We have developed a package of codes that can calculate dipole moment coefficients $\mathbf{\alpha}_{|s\rangle,|s'\rangle}$, transition matrix $\mathbf{M}_{T}$, and rate equations. These codes can be obtained through a public code deposit service \cite{code}.

After deriving the rate equations, we illustrate the evolution of atomic populations over time and the resulting efficiency $\beta$. 
Fig.~\ref{Fig3}(a) depicts the time dependent populations of states $5S_{1/2} |\Tilde{3},-\Tilde{3}\rangle$, $5S_{1/2} |\Tilde{3},-\Tilde{2}\rangle$, $5S_{1/2} |\Tilde{2},-\Tilde{2}\rangle$, and $5P_{3/2} |\Tilde{4}',-\Tilde{4}'\rangle$ at a magnetic field of 161~G, with $I_{img}$ and $I_{rep}$ set to $I_{sat}$ and $0.4 I_{sat}$. 
The imaging beam is polarized as $\sigma_{-}$, while the repumping beam is polarized as half $\sigma_{+}$ and half $\sigma_{-}$ . 
Initiallly, All atoms are prepared in the state $5S_{1/2}|\Tilde{2},-\Tilde{2}\rangle$.
The population of $5S_{1/2} |\Tilde{2},-\Tilde{2}\rangle$ drops to zero within 10~$\mathrm{\mu}$s, and the majority of atoms enter the closed cycle for imaging with an efficiency of approximately 75\%. However, some of the atoms enter other states, as shown in Fig.~\ref{Fig3}(a).
To further analyze the early transient behaviors, we introduce a settling time $\tau_{s}$, defined as the time for the decay of $f_{|\Tilde{2},-\Tilde{2}\rangle}$ from 1 to $1/e^2$. 
Once the imaging time exceeds $\tau_{s}$, the transient behaviors can be disregarded. 
In Fig.~\ref{Fig3}(b), we plot $\tau_{s}$ against $I_{rep}$, demonstrating that the settling time is less than 10 $\mathrm{\mu}$s when $I_{rep}$ exceeds 0.2$I_{sat}$ under 161~G magnetic field.
We also compare the results at 161~G and 0~G, revealing no order difference in the settling time with or without magnetic fields.
 
Fig.~\ref{Fig3}(c) displays the efficiency $\beta$ versus $I_{rep}$ with an imaging time of 40~$\mathrm{\mu s}$, indicating how the transient behaviors of population repumping affect $\beta$. We observe that $\beta$ changes $11\%$ within the intensity range of 0.1$I_{sat}$ to $I_{sat}$ due to the change of repumping efficiency.
When the repumping light is larger than $I_{sat}$, the change of $\beta$ is less than 1\%.
Fig.~\ref{Fig3}(d) displays the variation of $\beta$ with respect to the imaging intensity $I_{img}$. It can be observed that within the intensity range of 0.1$I_{sat}$ to 1.5$I_{sat}$, the change in $\beta$ is only 1.4\%. Therefore, it is a good approximation that the efficiency $\beta$ is independent on the imaging intensity. In most experiments, the imaging light is attenuated by atoms due to a large optical depth, resulting in different light intensities being experienced by each atom. However, the high repumping efficiency under a high magnetic field enables the imaging process with repumping transition $5S_{1/2}|\Tilde{2},-\Tilde{2}\rangle\rightarrow 5P_{1/2}|\Tilde{4}',-\Tilde{3}'\rangle$ to have a similar saturation intensity as the condition without a magnetic field. As shown in Fig.~\ref{Fig3}(d), the efficiency $\beta$ can be treated as approximately constant, at least at the leading order. Therefore, a universal $\beta$ can be used to describe the imaging process, provided that a precise measurement of the light intensity is not required.

\begin{figure}[bt]
\centering
\includegraphics[width=0.44\textwidth]{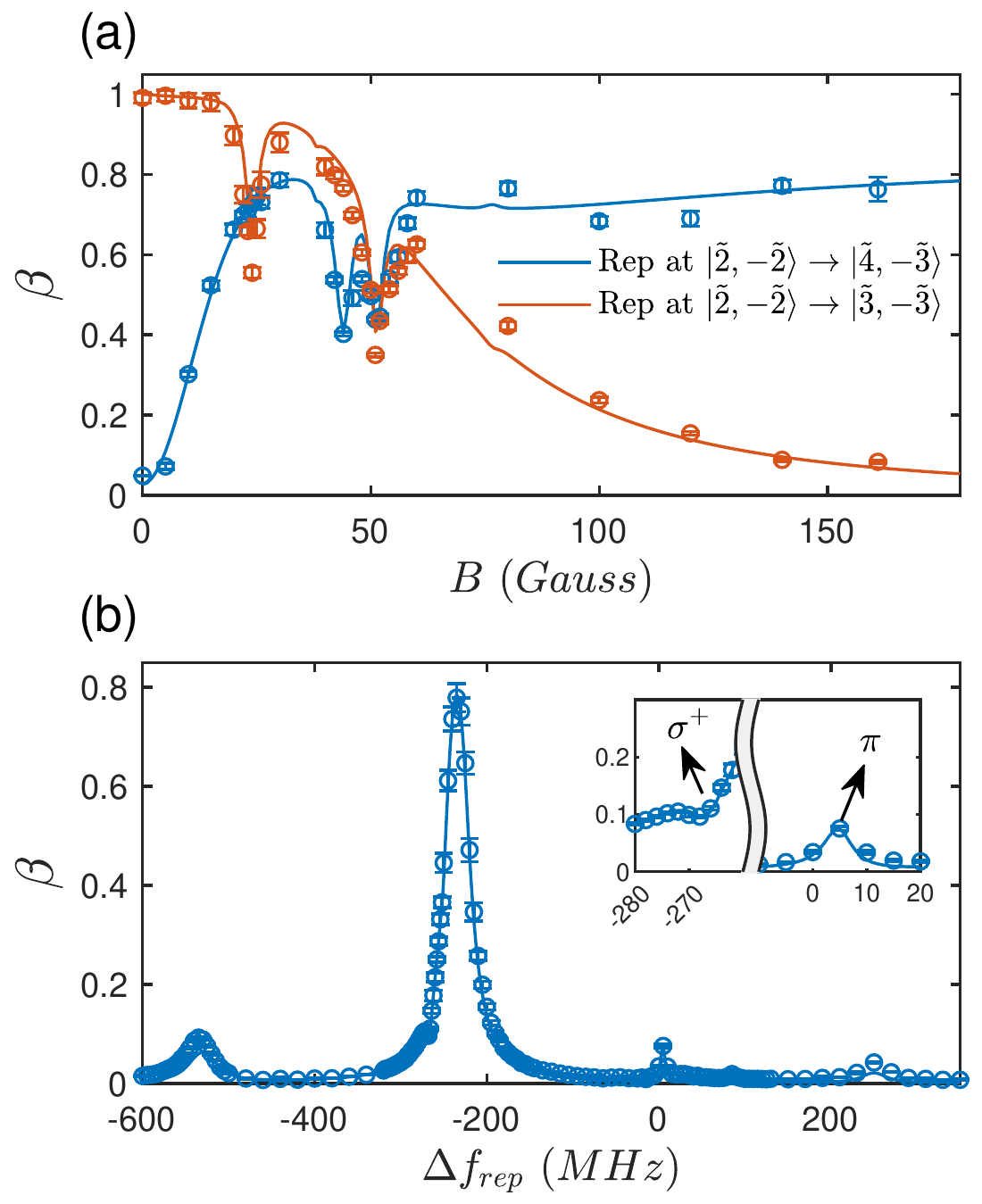}
\caption{(a) The imaging efficiency $\beta$ versus magnetic fields with two different repumping frequencies. The solid curves are theoretical calculations, and the scattered points are the experimental data while the error bars corresponds to one standard deviation. Each data point corresponds to the average of ten experimental measurements. The red curve and points correspond to the repumping transition $5S_{1/2}|\Tilde{2},-\Tilde{2}\rangle\rightarrow 5P_{3/2}|\Tilde{3}',-\Tilde{3}'\rangle$. The blue curve and points correspond to the repumping transition $5S_{1/2}|2,-2\rangle\rightarrow 5P_{3/2}|\Tilde{4}',-\Tilde{3}'\rangle$.
(b) $\beta$ versus the repumping frequency under 161~G magnetic field. 
Here the frequency zero point corresponds to the frequency of transition $5S_{1/2} |\Tilde{2},-\Tilde{2}\rangle\rightarrow 5P_{3/2} |\Tilde{3}',-\Tilde{3}'\rangle$ under 0~G.
There are two large peaks at -535~MHz and -235~MHz, corresponding to transition $5S_{1/2} |\Tilde{2},-\Tilde{2}\rangle\rightarrow 5P_{3/2} |\Tilde{3}',-\Tilde{3}'\rangle$ and transition $5S_{1/2} |\Tilde{2},-\Tilde{2}\rangle\rightarrow 5P_{3/2} |\Tilde{3}',-\Tilde{3}'\rangle$.
The inset shows more detail features around -270 and 0~MHz while there are peaks at 5~MHz and a valley at -268~MHz, which is attributed to the $\pi$ component and $\sigma^+$ component of the repumping light.
}\label{fig4}
\end{figure}

Fig.~\ref{fig4} shows the variation of $\beta$ with respect to the magnetic field ranging from 0 to 180~G. The solid curves represent the theoretical calculations, while the scattered points correspond to the experimental data. These curves will be further elaborated in the upcoming section, in conjunction with the experimental procedures. 
 
\section{Experimental verification}
Our experimental setup, depicted in Fig.~\ref{Fig1}(a), utilizes a crossed dipole trap with vibrational frequencies ($\omega_x$, $\omega_y$, $\omega_z$)/2$\pi$=(180, 180, 255)~Hz to confine 1.3$\times10^6$ rubidium-85 atoms. 
The atoms are initially prepared in the state $5S_{1/2}|2,-2\rangle$  which exhibits a broad Feshbach resonance at approximately 155~G. Prior to subsequent experiment, the atoms are held to reach thermal equilibrium at a temperature of 10~$\mu$K. The magnetic field is then adiabatically raised to the desired value and held for 400ms after which the dipole trap is turned off and the atoms are allowed to expand for the time-of-flight measurement. We then conduct absorption imaging at the target magnetic field using imaging light with $I_{img}=1.1I_{sat}$ and repumping light with $I_{rep}=0.6I_{sat}$. The imaging light is predominantly polarized in the $\sigma^-$ direction, while the repumping light with a linear polarization perpendicular to the magnetic field is split evenly between $\sigma^+$ and $\sigma^-$.
For comparison, we also need to perform the absorption imaging at zero magnetic field to determine the actual atom number. Due to different loss under different magnetic field, we also ramp up the magnetic field to the desired value, then hold atoms for the same time, and then ramp down the the magnetic field to zero for imaging. This ensures that the atoms experience the same loss process at both large magnetic fields and zero ones. 
 
By comparing these two results, we can get the efficiency $\beta$.

In Fig.~\ref{fig4}(a), we compare two cases with different repumping frequencies at desired magnetic fields. 
One is resonant with the transition $5S_{1/2}|\Tilde{2},-\Tilde{2}\rangle\rightarrow 5P_{3/2}|\Tilde{3}',-\Tilde{3}'\rangle$ (red curve and points) and another is resonant with the transition $5S_{1/2}|\Tilde{2},-\Tilde{2}\rangle\rightarrow 5P_{3/2}|\Tilde{4}',-\Tilde{3}'\rangle$ (blue curve and points). 
Under zero magnetic fields, the transition $5S_{1/2}|\Tilde{2},-\Tilde{2}\rangle\rightarrow 5P_{3/2}|\Tilde{4}',-\Tilde{3}'\rangle$ is dipole forbidden, so the efficiency $\beta$ is zero. As the magnetic field increases, the transition $5S_{1/2}|\Tilde{2},-\Tilde{2}\rangle\rightarrow 5P_{3/2}|\Tilde{4}',-\Tilde{3}'\rangle$ becomes increasingly allowed, resulting in a higher $\beta$.
Meanwhile, the $\sigma^+$ polarization of the repumping light, which would pump the atoms to levels further away from the closed transition, is supposed to decrease $\beta$.
Therefore, $\beta$ of the repumping frequency $5S_{1/2}|\Tilde{2},-\Tilde{2}\rangle \rightarrow 5P_{3/2}|\Tilde{4}',-\Tilde{3}'\rangle$ has a valley at 44~G due to the level cross of $5P_{3/2}|\Tilde{4}',-\Tilde{3}'\rangle$ and $5P_{3/2}|\Tilde{3}',-\Tilde{1}'\rangle$, and $\beta$ of the repumping frequency $ 5S_{1/2}|\Tilde{2},-\Tilde{2}\rangle \rightarrow  5P_{3/2}|\Tilde{3}',-\Tilde{3}'\rangle$ has a valley at 21~G due to the cross of $ 5P_{3/2}|\Tilde{3}',-\Tilde{3}'\rangle$ and $ 5P_{3/2}|\Tilde{2}',-\Tilde{1}'\rangle$. Similarly, both transitions exhibit a valley in their efficiency at 51~G where the energy of $5P_{3/2} |\Tilde{4}',-\Tilde{4}'\rangle$ and $5P_{3/2} |\Tilde{3}',-\Tilde{2}'\rangle$ cross, due to the residual $\sigma^+$ polarization in the imaging light.

In Fig.~\ref{fig4}(b), we show the efficiency $\beta$ versus the repumping light frequency at 161 G. 
The frequency zero point is defined as the transition $5S_{1/2} |\Tilde{2},-\Tilde{2}\rangle\rightarrow 5P_{3/2} |\Tilde{3}',-\Tilde{3}'\rangle$ under zero magnetic field.
The peaks at -530~MHz and -235~MHz correspond to the transitions $5S_{1/2} |\Tilde{2},-\Tilde{2}\rangle \rightarrow 5P_{3/2} |\Tilde{3}',-\Tilde{3}'\rangle$ and $5S_{1/2} |\Tilde{2},-\Tilde{2}\rangle \rightarrow 5P_{3/2} |\Tilde{4}',-\Tilde{3}'\rangle$, respectively, and are due to the $\sigma^-$ component of the repumping light. 
We can clearly see that the transition $5S_{1/2} |\Tilde{2},-\Tilde{2}\rangle \rightarrow 5P_{3/2} |\Tilde{4}',-\Tilde{3}'\rangle$ is the most efficient repumping frequency at high magnetic field and gives $\beta$ of 0.78. 
The peak at 250~MHz and the valley at -268~MHz correspond to the transitions $5S_{1/2} |\Tilde{2},-\Tilde{2}\rangle \rightarrow 5P_{3/2} |\Tilde{4}',-\Tilde{1}'\rangle$ and $5S_{1/2} |\Tilde{2},-\Tilde{2}\rangle \rightarrow 5P_{3/2} |\Tilde{2}',-\Tilde{1}'\rangle$, respectively, and are due to the $\sigma^+$ component of the repumping light. 
The small peak at 5~MHz corresponds to the transition $5S_{1/2} |\Tilde{2},-\Tilde{2}\rangle \rightarrow 5P_{3/2} |\Tilde{4}',-\Tilde{2}'\rangle$ and is due to the residual $\pi$ component of the repumping light.

Based on these structures, we fit the polarization impurity and obtain that the $\pi$ component of repumping light is 0.35\% which is consistent with our other independent polarization measurement at 0.2\% level. And the $\sigma^+$ component of imaging light is about 3\%. 
These features suggest that we should avoid any Zeeman level crossing during the absorption imaging. Also the imaging transition and the repumping transition should be carefully chosen to get best imaging efficiency.
\begin{figure}[bt]
\centering
\includegraphics[width=0.45\textwidth]{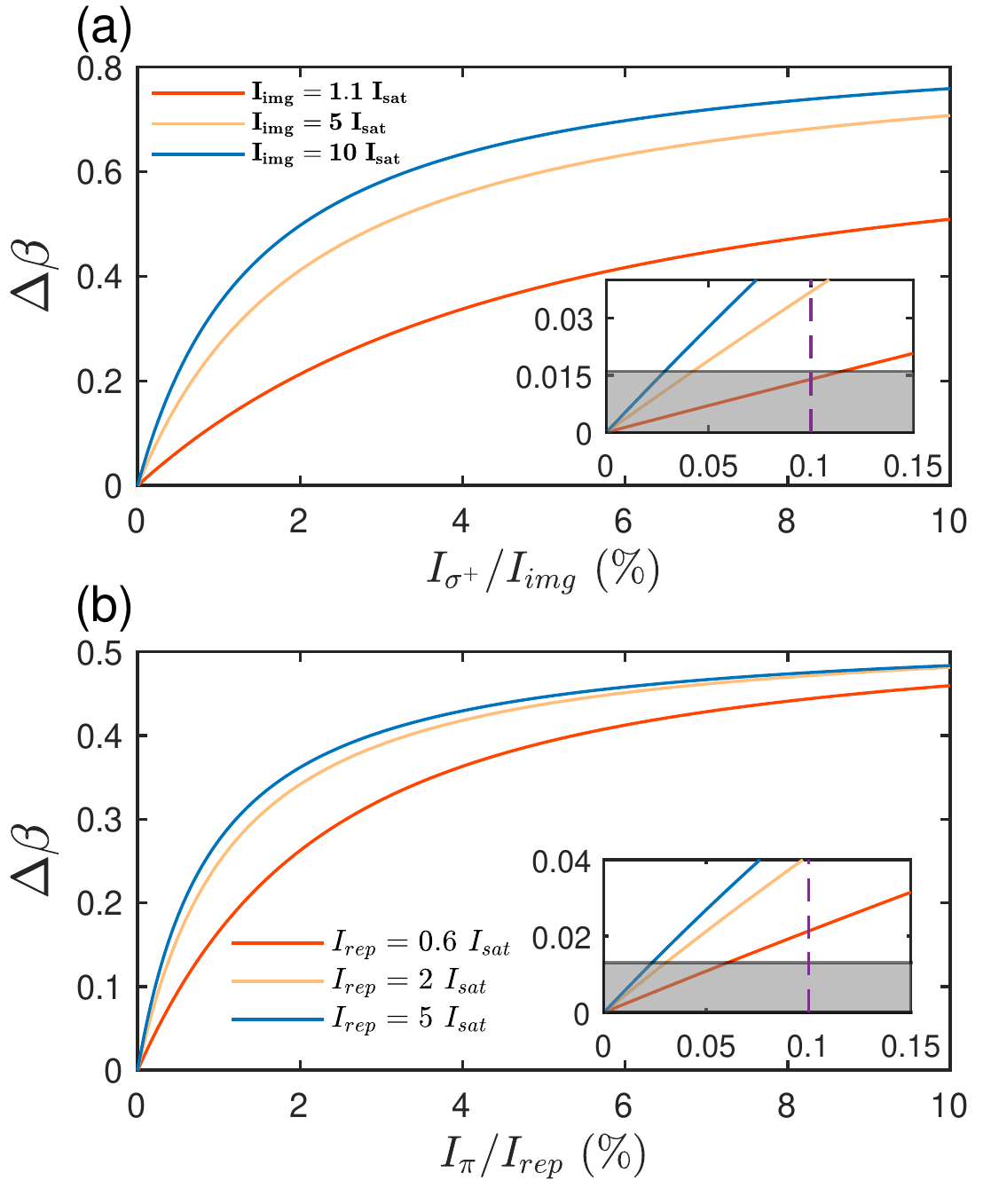}
\caption{(a) The variation of $\beta$ due to different fractions of $\sigma^+$ polarization in the imaging light at varying intensities of the imaging light. (b) The variation of $\beta$ due to different fractions of $\pi$ polarization in the repumping light at varying intensities of the repumping light. The shaded region in the inset represents the average error bars of our experimental measurements corresponding to the detection limits of $\Delta\beta$. The purple dashed line on the inset shows that the imperfection of the light is at the 0.1\% level, which is the performance achieved by commonly used polarization beam splitters such as Thorlabs PBS102.}\label{fig5}
\end{figure}

\section{APPLICATION}

In this section, we want to suggest a practical application of this high-field-imaging model for calibrating the laser beam polarization by atoms.

As shown in Fig.~\ref{fig4}, a slight $\sigma^+$ polarization in the imaging light can lead to a significant valley of $\beta$ at 51~G, while $\pi$ polarization in the repumping light can result in a peak at 161~G with repumping light at 5~MHz. Therefore, we can utilize the variation of $\beta$ to calibrate the polarization quality of both the imaging and repumping light.

We define $\Delta\beta$ as the change in $\beta$ caused by imperfections in the light sources, and show the corresponding curve in Figure~\ref{fig5}. Within the error bars of our experiment, we can detect imperfections as small as 0.04\% for $\sigma^+$ polarization in the imaging light with $I_{img}=5I_{sat}$ and 0.02\% for $\pi$ polarization in the repumping light with $I_{rep}=5I_{sat}$. To our knowledge, most of commercial polarizers such as Thorlabs PBS102 can only provide an extinction ratio of 1000:1 at the transmission port. Our measurement based on the spectrum can provide higher precision for calibrating the polarization of incident light.

\section{Conclusions}
To sum up, we have developed a theoretical model that accurately describes the absorption imaging of atoms in the presence of a magnetic field. Our model allows for the determination of a correction factor for atom number measurements and identification of optimal imaging parameters. Our experimental verification of the model shows that absorption spectroscopy provides a sensitive and useful tool for benchmarking laser beam quality, allowing for the detection of features such as polarization imperfections and Zeeman level crossings. This method will be helpful for experimental groups in calibrating absorption imaging and ensuring high-quality incident light.

\section*{Acknowledgement}
The code, calculating the dipole moments, rate equations, and imaging efficiency under magnetic fields, is available through the public code deposit service \cite{code}.

This work is financially supported by National Natural Science Foundation of China (61975092, 92165203, 11974202) and National Key Research and Development Program of China (2021YFA0718303, 2021YFA1400904) .

\bibliography{reference}

\end{document}